\documentclass[usenatbib]{emulateapj}
\usepackage{graphicx}
\usepackage[flushleft]{threeparttable}
\usepackage[usenames,dvipsnames,svgnames,table]{xcolor}
\usepackage{rotating}
\usepackage{hyperref}
\definecolor{darkblue}{rgb}{0.0,0.0,0.3}
\hypersetup{colorlinks,breaklinks,
            linkcolor=darkblue,urlcolor=darkblue,
            anchorcolor=darkblue,citecolor=darkblue}
\usepackage{amsmath,amssymb}
\usepackage{amsmath}

\begin{document}

\def\etal{et al.\ \rm}
\def\ba{\begin{eqnarray}}
\def\ea{\end{eqnarray}}
\def\etal{et al.\ \rm}
\def\Fdw{F_{\rm dw}}
\def\Tex{T_{\rm ex}}
\def\Fdis{F_{\rm dw,dis}}
\def\Fnu{F_\nu}
\def\FJ{F_J}
\def\rout{r_{\rm out}}

\newcommand\cmtrr[1]{{\color{red}[RR: #1]}}

%%%%%%%%%%%%%%%%%%%%%%%%%%%%%%%%%%%%%%%%%%%%%%%%%%%%%%%%%%%

\title{Protoplanetary Disks as (Possibly) Viscous Disks}

\author{Roman R. Rafikov\altaffilmark{1,2}}
\altaffiltext{1}{Department of Applied Mathematics and Theoretical Physics, Centre for Mathematical Sciences, University of Cambridge, Wilberforce Road, Cambridge CB3 0WA, UK; rrr@damtp.cam.ac.uk}
\altaffiltext{2}{Institute for Advanced Study, 1 Einstein Drive, Princeton NJ 08540}

%%%%%%%%%%%%%%%%%%%%%%%%%%%%%%%%%%%%%%%%%%%%%%%%%%%%%%%%%%%

\begin{abstract}
Protoplanetary disks are believed to evolve on Myr timescales in a diffusive (viscous) manner as a result of angular momentum transport driven by internal stresses. Here we use a sample of 26 protoplanetary disks resolved by {\it ALMA} with measured (dust-based) masses and stellar accretion rates to derive the dimensionless $\alpha$-viscosity values for individual objects, with the goal of constraining the angular momentum transport mechanism. We find that the inferred values of $\alpha$ do not cluster around a single value, but instead have a broad distribution extending from $10^{-4}$ to $0.04$. Moreover, they correlate with neither the global disk parameters (mass, size, surface density) nor the stellar characteristics (mass, luminosity, radius). However, we do find a strong linear correlation between $\alpha$ and the central mass accretion rate $\dot M$. This correlation is unlikely to result from the direct physical effect of $\dot M$ on disk viscosity on global scales. Instead, we suggest that it is caused by the decoupling of stellar $\dot M$ from the global disk characteristics in one of the following ways. (1) The behavior (and range) of $\alpha$ is controlled by a yet unidentified parameter (e.g. ionization fraction, magnetic field strength, or geometry), ultimately driving the variation of $\dot M$. (2) The central $\dot M$ is decoupled from the global viscous mass accretion rate as a result of an instability or mass accumulation (or loss) in the inner disk. (3) Perhaps the most intriguing possibility is that angular momentum in protoplanetary disks is transported non-viscously, e.g. via magnetohydrodynamic winds or spiral density waves. 
\end{abstract}


%%%%%%%%%%%%%%%%%%%%%%%%%%%%%%%%%%%%%%%%%%%%%%%%%%%%%%%%%%%

%%%%%%%%%%%%%%%%%%%%%%%%%%%%%%%%%%%%%%%%%%%%%%%%%%%%%%%%%%%
%%%%%%%%%%%%%%%%%%%%%%%%%%%%%%%%%%%%%%%%%%%%%%%%%%%%%%%%%%%

\section{Introduction}  
\label{sect:intro}

%%%%%%%%%%%%%%%%%%%%%%%%%%%%%%%%%%%%%%%%%%%%%%%%%%%%%%%%%%%

Protoplanetary disks are thought to persist around their parent stars for relatively short span of time. Observations present a clear evolutionary picture, in which both the fraction of systems exhibiting IR excess \citep{Hillenbrand} and the mass accretion rate onto the central star \citep{Calvet} decline of Myr timescales. Both observational indicators are thought to be the clear signatures of the presence of the circumstellar disks. 

Astrophysical accretion disks are believed to evolve predominantly under the action of their internal stresses \citep{shakura_1973,lynden-bell_1974}, and protoplanetary disks are no exception to the rule. For a long time evolutionary models of the protoplanetary disks have been developed assuming that the disks spread viscously, losing mass to the central star, while at the same time providing the birth site for planet formation. The characteristic time for the disk evolution in these models is simply the viscous time $t_\nu$ at the outer radius of the disk $r_{\rm out}$,
\ba
t_\nu\approx r_{\rm out}^2/\nu.
\label{eq:t_nu}
\ea
Here $\nu$ is the kinematic viscosity, which is conveniently parametrized using the $\alpha$-prescription \citep{shakura_1973}
\ba
\nu=\alpha c_s^2\Omega^{-1},
\label{eq:alpha}
\ea
with $\alpha\lesssim 1$ being constant, $c_s\equiv(kT/\mu)^{1/2}$ ($T$ is the disk temperature) and $\Omega\equiv (GM_\star/r^3)^{1/2}$ ($M_\star$ is the mass of the central star) being the local sound speed and Keplerian angular frequency. Viscous models invariably predict that on long timescales (exceeding the viscous time of the initial, more compact disk, so that $\rout$ grows beyond the initial disk radius) the central mass accretion rate $\dot M$ should be related to the total disk mass $M_d$ as
\ba
\dot M\approx M_d/t_\nu,
\label{eq:Mdot-Md}
\ea  
(up to a constant factor of order unity) with $t_\nu$ evaluated at $\rout$, see equation (\ref{eq:t_nu}).

The idea of the viscous evolution of the protoplanetary disks, diffusive in character and characterized by equations (\ref{eq:t_nu})-(\ref{eq:Mdot-Md}), has gained certain observational support. In particular, \citet{Hartmann} and \citet{Calvet} found that the observed average properties of protoplanetary disks can be explained by their viscous evolution with efficiency corresponding to $\alpha\approx 10^{-2}$. 

The value of the dimensionless parameter $\alpha$ is believed to directly reflect the physics of the mechanism responsible for the angular momentum transport in the disk. In hot and well ionized accretion disks around compact objects transport is generally thought to be mediated by the magneto-rotational instability (MRI; \citealt{Velikhov,Chandrasekhar,Balbus}). Situation is less clear in the cold and poorly ionized protoplanetary disks, where the non-ideal magnetohydrodynamic (MHD) effects are known to weaken or even suppress the transport driven by the MRI \citep{Turner}. Other potential candidates such as gravitoturbulence \citep{Gammie,Rafikov15}, Rossby-wave instabilitiy \citep{Lovelace}, convective over-stability \citep{Latter}, vertical shear instability \citep{Urpin,Stoll}, and so on, have been proposed to explain the observed evolution of the protoplanetary disks properties. 

On the other hand, recent studies argue that the non-MRI related transport mechanisms can hardly be responsible for the observed disk evolution on the Myr timescales, because of the weakness of the purely hydrodynamic transport mechanisms \citep{Stoll,Turner}. Partly for that reason, the {\it non-diffusive} angular momentum transport mechanisms such as MHD winds \citep{Wardle,Suzuki,Bai} or spiral shocks \citep{Rafikov02,Rafikov16} have been gaining popularity. The distinctive feature of these mechanisms is that they do not need to obey the equations (\ref{eq:t_nu})-(\ref{eq:Mdot-Md}), thus resulting in a different relation between $\dot M$ and $M_d$.

The advent of {\it ALMA} made possible more precise and focused efforts to understand protoplanetary disk evolution. Recent measurements of the continuum and CO line emission for a large sample of protoplanetary disks in Lupus by \citet{Ansdell} and \citet{Miotello}, coupled with the most up-to-date determinations of the mass accretion rate $\dot M$ onto their parent stars by \citet{Alcala14,Alcala16}, allowed \citet{Manara} to identify a correlation between the disk mass $M_d$ and the central mass accretion rate. The disk masses were derived using the dust masses inferred from the continuum sub-mm emission assuming fixed gas-to-dust ratio. This correlation has been interpreted by \citet{Manara} as providing evidence for the viscous character of the protoplanetary disk evolution, in which the global disk properties directly determine the mass accretion rate at its center. 

In this work we focus on a different diagnostics of the viscous disk evolution. Using a sample of protoplanetary disks {\it directly resolved} by {\it ALMA}, with measured dust and gas masses \citep{Ansdell}, as well as the central accretion rates \citep{Alcala14,Alcala16}, we provide a {\it direct determination} of the value of the $\alpha$-parameter in individual systems. Given that different mechanisms of the angular momentum transport in disks predict different values of $\alpha$, our effort can provide direct information on the physical nature of the internal stresses driving the disk evolution. Unlike other studies \citep{Hartmann,Jones}, in this work we (1) utilize information about the individual disk sizes provided by {\it ALMA} and (2) do not use information on the ages of the parent stars, which are known to be very uncertain. 

Our work is organized as follows. We describe our methodology for inferring $\alpha$ in \S \ref{sect:method}, and our observational sample in \S \ref{sect:sample}. Our results, including correlations of $\alpha$ with different characteristics of the observed systems, can be found in \S \ref{sect:results}. We provide extensive discussion of our findings in \S \ref{sect:disc}, and summarize the results in \S \ref{sect:sum}.

%%%%%%%%%%%%%%%%%%%%%%%%%%%%%%%%%%%%%%%%%%%%%%%%%%%%%%%%%%%
%%%%%%%%%%%%%%%%%%%%%%%%%%%%%%%%%%%%%%%%%%%%%%%%%%%%%%%%%%%

\section{Methodology}  
\label{sect:method}

%%%%%%%%%%%%%%%%%%%%%%%%%%%%%%%%%%%%%%%%%%%%%%%%%%%%%%%%%%%

In our analysis we will assume that, as a result of expansion driven by internal stresses, the present day sizes of the protoplanetary disks in Lupus exceed their initial radii, set at the mass infall phase. Then the disk can be approximately considered as evolving in a self-similar fashion, and equations (\ref{eq:t_nu})-(\ref{eq:Mdot-Md}) should apply. Their combination yields
\ba
\alpha\approx \frac{\dot M}{M_d}\frac{\mu}{kT}\Omega r^2,
\label{eq:alpha_expr}
\ea
where $\Omega$, $r$, $T$ are evaluated at the outer disk radius $r_{\rm out}$, and $M_d$ is the disk mass enclosed within $r_{\rm out}$. 

In equation (\ref{eq:alpha_expr}) the values of $r_{\rm out}$, $M_d$, $\dot M$, and $M_\star$ come directly from observations. However, to obtain $\alpha$ we still need to make assumptions about the disk temperature $T(r_{\rm out})$. We try three different thermodynamic prescriptions in this work.

First, we simply assume that
\ba
T(r_{\rm out})=20~\mbox{K},
\label{eq:T_fixed}
\ea
for all disks in our sample. This prescription is the same as that used by \citet{Ansdell} for deriving the dust masses $M_{\rm dust}$ of the disks from their continuum sub-mm fluxes, providing certain internal consistency.

%%%%%%%%%%%%%%%%%%%%%%%%%%%
\begin{figure}
\centering
\includegraphics[width=0.5\textwidth]{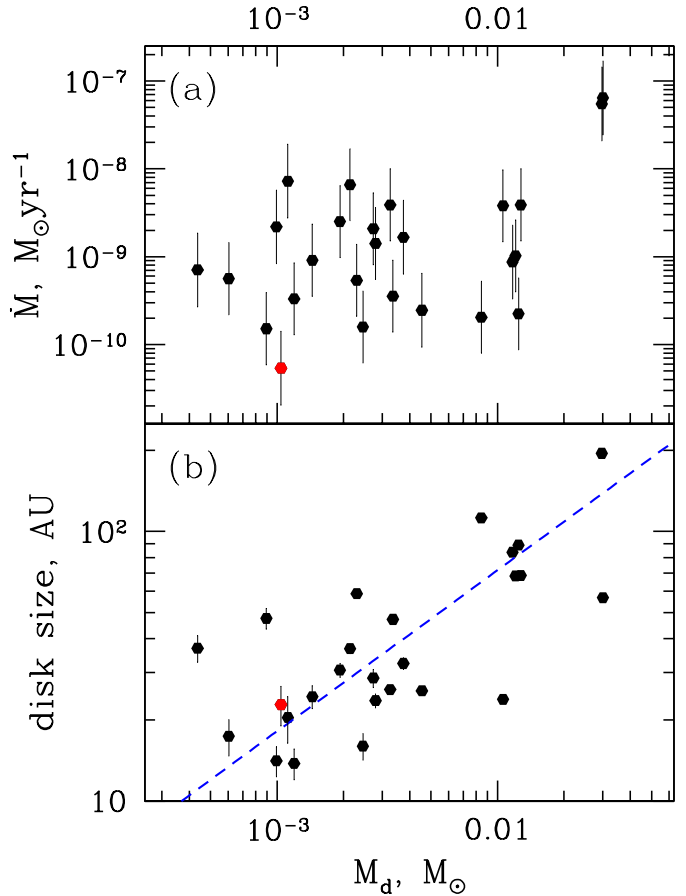}
\caption{
(a) Central accretion rate $\dot M$ and (b) disk size $r_{\rm out}$ for the sample of objects used in this work (Table \ref{table:sample}), plotted vs. the disk mass $M_d$ inferred from the continuum dust emission. While we do not find strong evidence for a correlation between $\dot M$ and $M_d$ in our sample [cf. \citet{Manara}], we do observe a correlation between $r_{\rm out}$ and $M_d$, with the linear regression shown as the blue dashed line. Red dot shows object  2MASS-J16081497-3857145, which is close to the brown dwarf regime. See text for more details. 
\label{fig:sample}}
\end{figure}
%%%%%%%%%%%%%%%%%%%%%%%%%%%

Second, we take the $T(r_{\rm out})$ to correspond to the temperature of optically thin dust, directly illuminated by the central star, in which case
\ba    
T(r_{\rm out})=\left(\frac{L_\star}{16\pi \sigma r_{\rm out}^2}\right)^{1/4}.
\label{eq:T_thin}
\ea  
This expression neglects the difference between the emission and absorption efficiencies of the grains. Stellar luminosity $L_\star$ is known to us from observations.

Finally, we also use a prescription for the optically thick, externally irradiated passive protoplanetary disks, motivated by \citet{Chiang}, that reads
\ba   
T(r_{\rm out})=120~\mbox{K}\left(\frac{L_\star}{L_\odot}\right)^{2/7}\left(\frac{M_\odot}{M_\star}\right)^{1/7}\left(\frac{\mbox{AU}}{r_{\rm out}}\right)^{3/7}.
\label{eq:T_thick}
\ea    
This prescription predicts $T(\rout)$ lower than in the case (\ref{eq:T_thin}).

We determine the full disk masses using the dust masses $M_{\rm dust}$ derived from the continuum sub-mm fluxes \citep{Ansdell}. To convert $M_{\rm dust}$ to the full disk mass $M_d$ we use a uniform gas-to-dust mass ratio $\chi=100$. In certain cases (\S \ref{sect:mdot}) we also use the information on the gas masses coming from the $^{13}$CO and C$^{18}$O line measurements by {\it ALMA}. However, it should be kept in mind that the disk masses inferred this way are believed to be systematically underestimated, often by more than an order of magnitude, as a result CO freeze out on dust grains or sequestration of carbon into large bodies \citep{Ansdell,Miotello16}. As a result, it is expected that $M_d=\chi M_{\rm dust}$, which we employ in this work, should provide a better estimate of the disk mass.

%%%%%%%%%%%%%%%%%%%%%%%%%%%%%%%%%%%%%%%%%%%%%%%%%%%%%%%%%%%

\subsection{Observational sample}  
\label{sect:sample}

%%%%%%%%%%%%%%%%%%%%%%%%%%%%%%%%%%%%%%%%%%%%%%%%%%%%%%%%%%%

Our approach to determining $\alpha$ via the equation (\ref{eq:alpha_expr}) works only for disks that have non-trivial measurements of $\dot M$, $M_d$, and $r_{\rm out}$, as well as of $L_\star$ and $M_\star$. For this reason, we are interested only in {\it resolved disks} with significant detections of both the continuum dust emission by {\it ALMA} and the stellar mass accretion rate $\dot M$ via spectroscopy.

\citet{Ansdell} have carried out {\it ALMA} survey of 89 protoplanetary disks in Lupus star-forming complex at $\sim 150-200$ pc away from the Sun (age $3\pm 2$ Myr, \citealt{Alcala14}). They directly resolved many sources and provided initial measurements of the dust and gas masses for about 2/3 and 1/3 of their sample, correspondingly. \citet{Miotello} carried out a more sophisticated analysis of this data set based on work of \citet{Miotello16}, providing more accurate dust and gas mass measurements. At the same time, \citet{Alcala14,Alcala16} carried out X-shooter spectroscopy for many of these targets, deriving central mass accretion rates $\dot M$ based on the UV excesses.

By examining the samples presented in these studies we selected 26 objects, which possess resolved disks with well measured $M_{\rm dust}$ and $\dot M$. Out of these disks 18 also have significant measurements of the gas mass based on CO line measurements. Two disks --- Sz84 and MYLup --- fall in the transitional disk category \citep{Alcala16}. The parameters of all 26 system are listed in Table \ref{table:sample}. We adopt $M_{\rm dust}$ and disk sizes from \citet{Ansdell}, gas masses $M_g$ from \citet{Miotello}, and $\dot M$ and stellar parameters from \citet{Alcala14,Alcala16}. 

For simplicity, in this study we associate the outer disk radius $r_{\rm out}$ with the semi-major axis obtained in \citet{Ansdell} by simple Gaussian fitting of the resolved continuum intensity pattern. This alone may introduce a systematic duncertainty in the determination of $\rout$ at the level of tens of per cent. Even more serious error may arise from the possible difference between the radii of the gas and dust disks, evidence for which has been found in a number of systems \citep{Andrews,Cleeves,Walsh}. We discuss the impact of the $\rout$ uncertainty on our results in \S \ref{sect:factors}. 

We show some characteristics of our systems in Figure \ref{fig:sample}. We single out one object --- 2MASS-J16081497-3857145 --- which is close to the brown dwarf regime and is different from the rest of the sample (shown as a red dot). 

%%%%%%%%%%%%%%%%%%%%%%%%%%%
\begin{figure}
\centering
\includegraphics[width=0.5\textwidth]{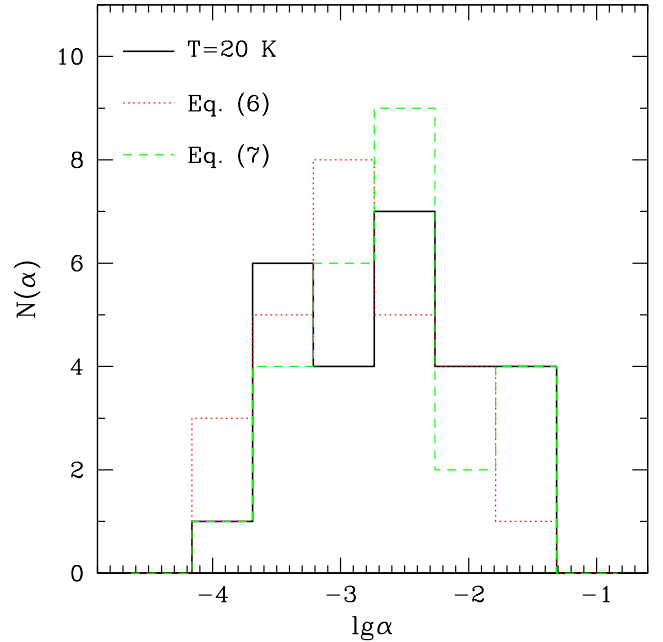}
\caption{
Distribution of the inferred values of $\alpha$, computed via equation (\ref{eq:alpha_expr}). Solid black, dotted red, and dashed green histograms correspond to $T(r_{\rm out})$ prescriptions given by the equations (\ref{eq:T_fixed}), (\ref{eq:T_thin}), and (\ref{eq:T_thick}), correspondingly. One can see a large spread in the values of $\alpha$, covering more than two orders of magnitude.
\label{fig:hist}}
\end{figure}
%%%%%%%%%%%%%%%%%%%%%%%%%%%

The top panel shows that, unlike \citet{Manara}, we do not observe a significant correlation between $M_d$ and $\dot M$ (see Table \ref{table:stats} for the statistical parameters of correlations shown in our plots: Pearson correlation coefficient $\rho$, Spearman's rank correlation coefficient, $r_s$, and $p$-value --- probability of the null hypothesis that the two variables have zero correlation). This is most likely explained by the modest size of our sample compared to that of \citet{Manara}. 

%%%%%%%%%%%%%%%%%%%%%%%%%%%
\begin{figure*}
\centering
\includegraphics[width=1\textwidth]{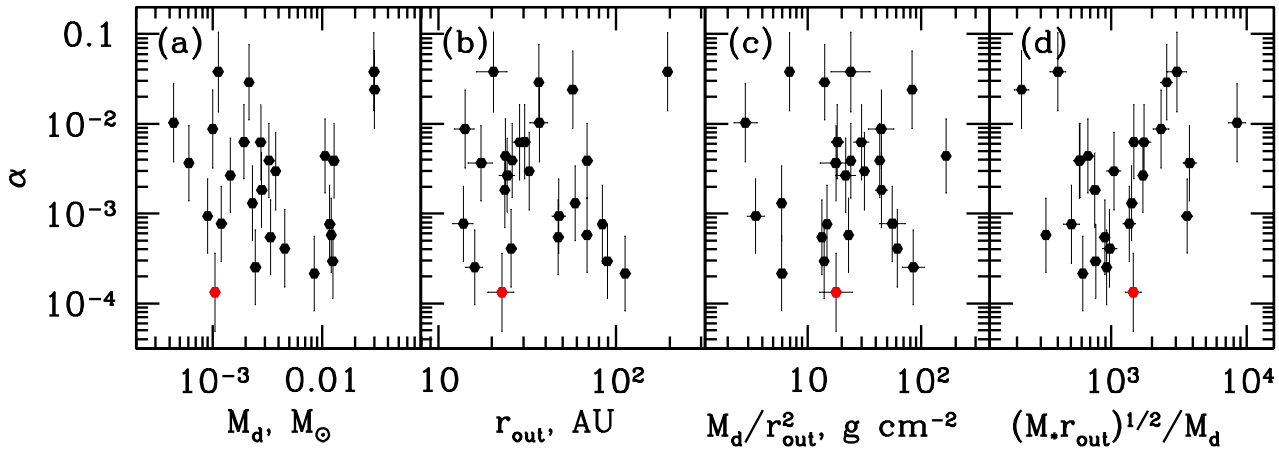}
\caption{Values of $\alpha$ plotted against some global characteristics of the disk: (a) disk mass $M_d$, (b) outer radius $r_{\rm out}$, (c) $M_d/r_{\rm out}^2$, which is a proxy for the surface density at $r_{\rm out}$, and (d) $(M_\star\rout)^{1/2}/M_d$, which is a combination of variables entering equation (\ref{eq:alpha_expr}). No statistically significant correlations between $\alpha$ and these global variables are found (see Table \ref{table:stats} for quantitative metrics).
\label{fig:disk_corrs}}
\end{figure*}
%%%%%%%%%%%%%%%%%%%%%%%%%%%

On the other hand, we do find an appreciable correlation between the disk size and mass, as Figure \ref{fig:sample}b demonstrates --- more extended disks typically have larger dust masses. The best fit bisector regression \citep{Isobe} describing this correlation is 
$\lg r_{\rm out}=(3.05 \pm 0.16)+(0.6 \pm 0.06)\lg \dot M$ (with $\rout$ and $\dot M$ measured in AU and $M_\odot$ yr$^{-1}$), but there is significant scatter around it. This relation may seem to suggest that the values of the disk surface density at the outer edge $\Sigma(r_{\rm out})\propto M_d/r_{\rm out}^2$ should be roughly the same. This could raise a worry that $\rout$, interpreted by \citet{Ansdell} as the outer extent of the disk, in fact corresponds to the detection limit of {\it ALMA}. However, Figure \ref{fig:disk_corrs}c shows that this is not the case, and that $M_d/r_{\rm out}^2$ for the observed sample spans almost two orders of magnitude, thanks to the large scatter in Figure \ref{fig:sample}b.

%%%%%%%%%%%%%%%%%%%%%%%%%%%%%%%%%%%%%%%%%%%%%%%%%%%%%%%%%%%
%%%%%%%%%%%%%%%%%%%%%%%%%%%%%%%%%%%%%%%%%%%%%%%%%%%%%%%%%%%

\section{Results}  
\label{sect:results}

%%%%%%%%%%%%%%%%%%%%%%%%%%%%%%%%%%%%%%%%%%%%%%%%%%%%%%%%%%%

In Figure \ref{fig:hist} we show the histograms for the values of $\alpha$ computed through equation (\ref{eq:alpha_expr}) for different thermodynamic assumptions, as shown on the panel. One can see that different methods of calculating the outer disk temperature do not result in large differences in the values of $\alpha$. Regardless of our assumptions, the distribution of $\alpha$ does not seem to show complicated substructure, roughly consistent with either being peaked (for $T(r_{\rm out})$ given by equations (\ref{eq:T_thin}) and (\ref{eq:T_thick})) or approximately uniform (for $T(r_{\rm out})=20$ K).

The most important feature of these distributions is their broad range. Irrespective of the $T(r_{\rm out})$ prescription, we find that in our sample of 26 disks the values of $\alpha$ span more than two orders of magnitude --- from $10^{-4}$ to $0.04$. This spread is hardly compatible with the simple idea of a single angular momentum transport mechanism setting the value of $\alpha$, as one would then expect a narrowly peaked distribution of $\alpha$ values. Nor could it be several physical mechanisms operating in different systems (e.g. different instabilities driving the transport), as then one would expect to see more substructure in the distribution of $\alpha$.

%%%%%%%%%%%%%%%%%%%%%%%%%%%
\begin{figure}
\centering
\includegraphics[width=0.5\textwidth]{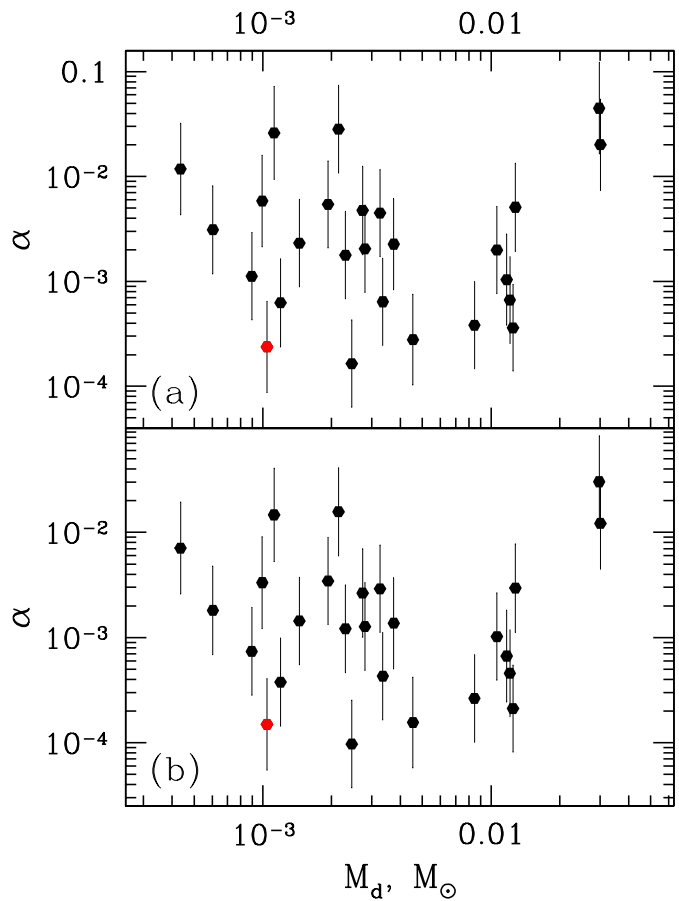}
\caption{Values of $\alpha$ computed assuming $T(\rout)$ given by (a) equation (\ref{eq:T_thin}) and (b) equation (\ref{eq:T_thick}), plotted as a function of the disk mass $M_d$. No correlations emerge here, in agreement with Figure \ref{fig:disk_corrs}a.
\label{fig:Mdisk_alpha_T}}
\end{figure}
%%%%%%%%%%%%%%%%%%%%%%%%%%%

As we show in \S \ref{sect:factors}, this spread is unlikely to be forced by the intrinsic scatter or observational errors in our sample. Thus, we are left to hypothesize that there may be some real physical reasons for this behavior of $\alpha$ in different systems, and we try to identify them next.

%%%%%%%%%%%%%%%%%%%%%%%%%%%%%%%%%%%%%%%%%%%%%%%%%%%%%%%%%%%

\subsection{Dependence of $\alpha$ on global disk properties} \label{sect:global}

%%%%%%%%%%%%%%%%%%%%%%%%%%%%%%%%%%%%%%%%%%%%%%%%%%%%%%%%%%%

What we are calculating via equation (\ref{eq:alpha_expr}), given the observables, is the value of $\alpha$ at $r_{\rm out}$, which determines the {\it global} evolution of the disk. For this reason it is natural to seek possible connection of $\alpha$ with the global variables characterizing disk on scales $\sim r_{\rm out}$. 

In Figure \ref{fig:disk_corrs} we plot $\alpha$ computed for $T(r_{\rm out})=20$ K versus the disk mass $M_d$, its radial extent $r_{\rm out}$, $M_d/r_{\rm out}^2$, which characterizes the surface density at $r_{\rm out}$, and $(M_\star\rout)^{1/2}M_d^{-1}$, which appears in equation (\ref{eq:alpha_expr}) together with $\dot M$. The errors on $\alpha$ were calculated quadratically from the uncertainties of the observables, as follows from equation (\ref{eq:alpha_expr}). It is clear that these plots do not reveal significant correlations of $\alpha$ with these global variables. This is also confirmed by the quantitative metrics of the possible relations between the pairs of variables shown in Table \ref{table:stats}. 

One may wonder whether this lack of correlation with the global disk parameters is forced by our simple assumption about the thermal state of the disk, represented by the equation (\ref{eq:T_fixed}). To address this issue, in Figure \ref{fig:Mdisk_alpha_T} we show the analog of Figure \ref{fig:disk_corrs}a, i.e. $\alpha$ vs. $M_d$, but calculated for $T(r_{\rm out})$ given by equations (\ref{eq:T_thin}) and (\ref{eq:T_thick}). One can see that, again, no correlation is present in the data, implying that this result is robust with regard to our assumptions about the disk temperature structure. To summarize, we do not find any clear dependence of $\alpha$ on the most obvious global characteristics of the disk. 

%%%%%%%%%%%%%%%%%%%%%%%%%%%
\begin{figure}
\centering
\includegraphics[width=0.5\textwidth]{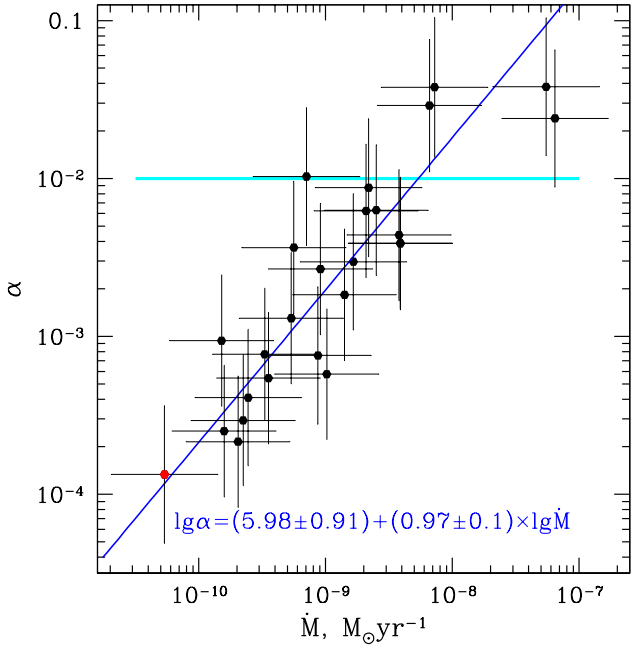}
\caption{Effective viscosity $\alpha$, computed for $T=20$ K (equation \ref{eq:T_fixed}), as a function of the central mass accretion rate $\dot M$. One can clearly see a strong correlation between the two variables. Blue line is the best fit to the data given by equation (\ref{eq:alpha_Mdot}). Horizontal cyan line is the dependence, around which the data points would be expected to cluster if the angular momentum transport were characterized by a single value of $\alpha$ (taken to be equal to $10^{-2}$ for illustrative purposes). Such clustering is clearly not exhibited by our sample, necessitating modifications to the simple picture of the viscous evolution of the protoplanetary disks.
\label{fig:Mdot_alpha}}
\end{figure}
%%%%%%%%%%%%%%%%%%%%%%%%%%%

%%%%%%%%%%%%%%%%%%%%%%%%%%%%%%%%%%%%%%%%%%%%%%%%%%%%%%%%%%%

\subsection{Dependence of $\alpha$ on $\dot M$}  
\label{sect:mdot}

%%%%%%%%%%%%%%%%%%%%%%%%%%%%%%%%%%%%%%%%%%%%%%%%%%%%%%%%%%%

Effective viscosity computed via equation (\ref{eq:alpha_expr}) depends not only on the global disk characteristics, but also on the central mass accretion rate $\dot M$. In Figure \ref{fig:Mdot_alpha} we plot the effective viscosity $\alpha$ for $T(r_{\rm out})=20$ K vs. $\dot M$. This Figure clearly reveals a strong correlation (Pearson coefficient $\rho(\alpha,\dot M)=0.877$) between $\alpha$ and $\dot M$. Simple linear bisector regression \citep{Isobe} results in a best fit line 
\ba
\lg \alpha=(5.98\pm 0.91)+(0.97\pm 0.1)\lg\dot M,
\label{eq:alpha_Mdot}
\ea
(with $\dot M$ measured in $M_\odot$ yr$^{-1}$) which is consistent with a linear dependence. This relation links the broad distribution of $\alpha$ seen in Figure \ref{fig:hist} with the spread of $\dot M$ in our sample. 

%%%%%%%%%%%%%%%%%%%%%%%%%%%
\begin{figure}
\centering
\includegraphics[width=0.5\textwidth]{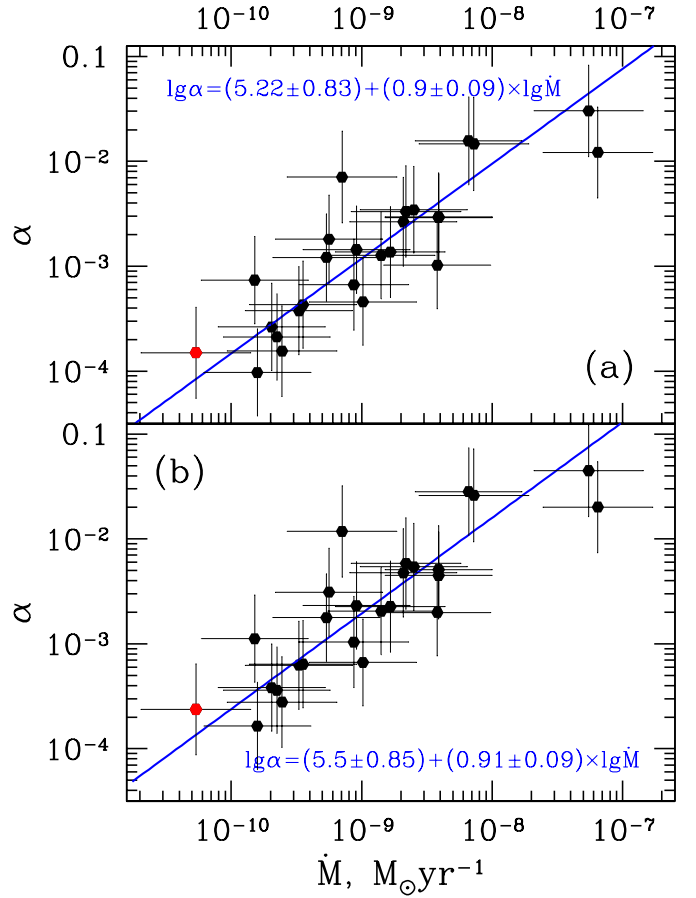}
\caption{Correlations between the central mass accretion rate $\dot M$ and $\alpha$, computed using different assumptions about disk temperature: (a) $T(r)$ given by equation (\ref{eq:T_thin}), and (b) $T(r)$ given by equation (\ref{eq:T_thick}). Correlation between $\dot M$ and $\alpha$ clearly persists in both cases. 
\label{fig:Mdot_alpha_T}}
\end{figure}
%%%%%%%%%%%%%%%%%%%%%%%%%%%

The $\alpha$-$\dot M$ correlation is robust with respect to our assumptions about $T(r_{\rm out})$, as further demonstrated in Figure \ref{fig:Mdot_alpha_T}. There we again observe that $\dot M$ and the effective viscosity parameter computed using equations (\ref{eq:T_thin}) and (\ref{eq:T_thick}) are strongly correlated, despite the different assumptions about disk thermodynamics. 

%%%%%%%%%%%%%%%%%%%%%%%%%%%
\begin{figure}
\centering
\includegraphics[width=0.5\textwidth]{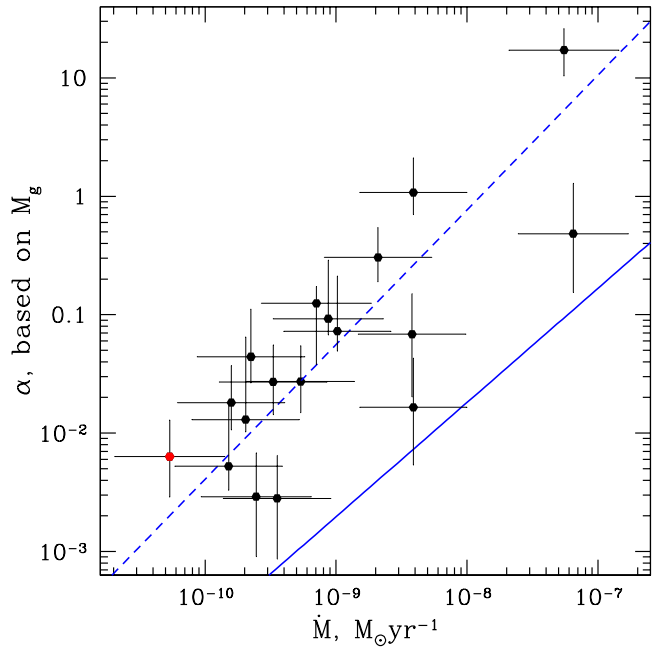}
\caption{
$\alpha$ computed using gas masses $M_g$ inferred from the CO line observations, instead of the dust-based masses $M_d$ computed using the continuum sub-mm emission (displayed in Figures \ref{fig:Mdot_alpha}-\ref{fig:Mdot_alpha_T}), shown as a function of $\dot M$. Despite the use of a different tracer of the disk mass, the $\alpha$-$\dot M$ correlation (dahed line) is still present at high significance. Solid line corresponds to the correlation (\ref{eq:alpha_Mdot}), which is clearly offset from the best fit for CO-based $\alpha$ and $\dot M$. 
\label{fig:alpha_gas}}
\end{figure}
%%%%%%%%%%%%%%%%%%%%%%%%%%%

%%%%%%%%%%%%%%%%%%%%%%%%%%%
\begin{figure*}
\centering
\includegraphics[width=1\textwidth]{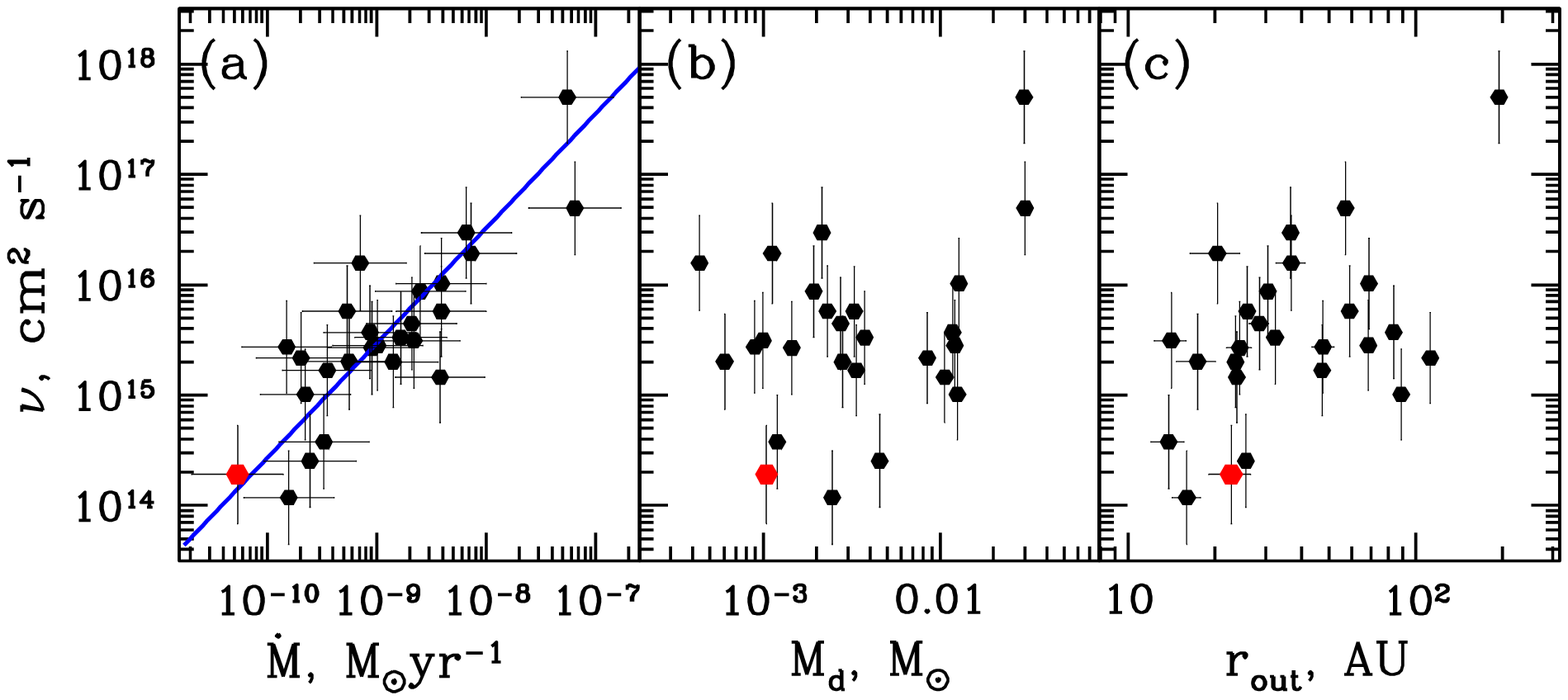}
\caption{Dimensional kinematic viscosity $\nu$ characterizing global disk evolution, computed for $T=20$ K, as a function of several global parameters: (a) central mass accretion rate $\dot M$, (b) disk mass $M_d$, and (c) the outer disk radius $\rout$. There is a clear correlation between $\nu$ and $\dot M$, but no correlation between $\nu$ and either $M_d$ or $\rout$.
\label{fig:disk_corrs_nu}}
\end{figure*}
%%%%%%%%%%%%%%%%%%%%%%%%%%%

Correlation persists even if use the CO-based gas masses $M_g$, available for 18 objects in our sample, instead of the dust-based masses $M_d$ when computing $\alpha$. This is illustrated in Figure \ref{fig:alpha_gas}. The spread in the values of $\alpha$ measured this way is considerably larger than in Figures \ref{fig:Mdot_alpha}-\ref{fig:Mdot_alpha_T}, and the best fit line is significantly offset from the relation (\ref{eq:alpha_Mdot}), illustrating the problem with the CO-based disk masses \citep{Ansdell,Miotello16}. 

If we were to take the CO-based masses at face value, we would conclude from Figure \ref{fig:alpha_gas} that, in the framework of the viscous model for the disk evolution based on equations (\ref{eq:t_nu})-(\ref{eq:Mdot-Md}), some systems require $\alpha\gtrsim 1$. As such values of $\alpha$ are unlikely, this could, again, demonstrate the problem with the disk mass determinations based on the CO line emission. 

The existence of a tight correlation between $\alpha$ and $\dot M$ is a non-trivial and rather unexpected result. Indeed, if the angular momentum transport in the disk were effected by a mechanism characterized by a unique value of $\alpha$, then Figure \ref{fig:Mdot_alpha} would look very differently, with $\alpha$ clustering around a well-defined value regardless of $\dot M$, and the slope of $\alpha(\dot M)$ relation being close to zero, as illustrated by the cyan line in this Figure. The $\alpha=10^{-2}$ corresponding to this line is for illustrative purposes, although this value has been suggested by the past studies \citep{Hartmann}. In that case the variation of $\dot M$ would have been exactly compensated by the variation of $\Omega r_{\rm out}^2/(M_d T)$. Figure \ref{fig:disk_corrs}d shows that the latter variable (proportional to $(M_\star\rout)^{1/2}M_d^{-1}$) exhibits essentially no correlation with $\alpha$, unlike $\dot M$ entering the expression (\ref{eq:alpha_expr}) for $\alpha$ in an identical fashion. Thus, the very fact that a strong $\alpha$-$\dot M$ correlation exists tells us something interesting.  

We also explored the behavior of the dimensional kinematic viscosity $\nu\approx r^2\dot M/M_d$, which plays the role of a diffusion coefficient for viscous spreading of the disk, see Figure \ref{fig:disk_corrs_nu}. One possible advantage of using $\nu$ instead of $\alpha$ is that its determination does not involve assumptions about the thermodynamic properties of the disk. One can see that $\nu$ is also strongly correlated with $\dot M$ (although the spread around the best fit line is larger than in Figures \ref{fig:Mdot_alpha}-\ref{fig:Mdot_alpha_T}), while at the same time being independent of either $M_d$ or $\rout$. This again suggests that there is a certain causal relation between $\dot M$ and the inferred disk viscosity.

%%%%%%%%%%%%%%%%%%%%%%%%%%%%%%%%%%%%%%%%%%%%%%%%%%%%%%%%%%%

\subsection{Dependence of $\alpha$ on stellar properties}  
\label{sect:stellar}

%%%%%%%%%%%%%%%%%%%%%%%%%%%%%%%%%%%%%%%%%%%%%%%%%%%%%%%%%%%

Having found correlation of $\alpha$ with $\dot M$, which is a {\it local} characteristic measured at the star, we also checked if $\alpha$ could have some relation to other  stellar parameters.

\begin{figure*}
\centering
\includegraphics[width=1\textwidth]{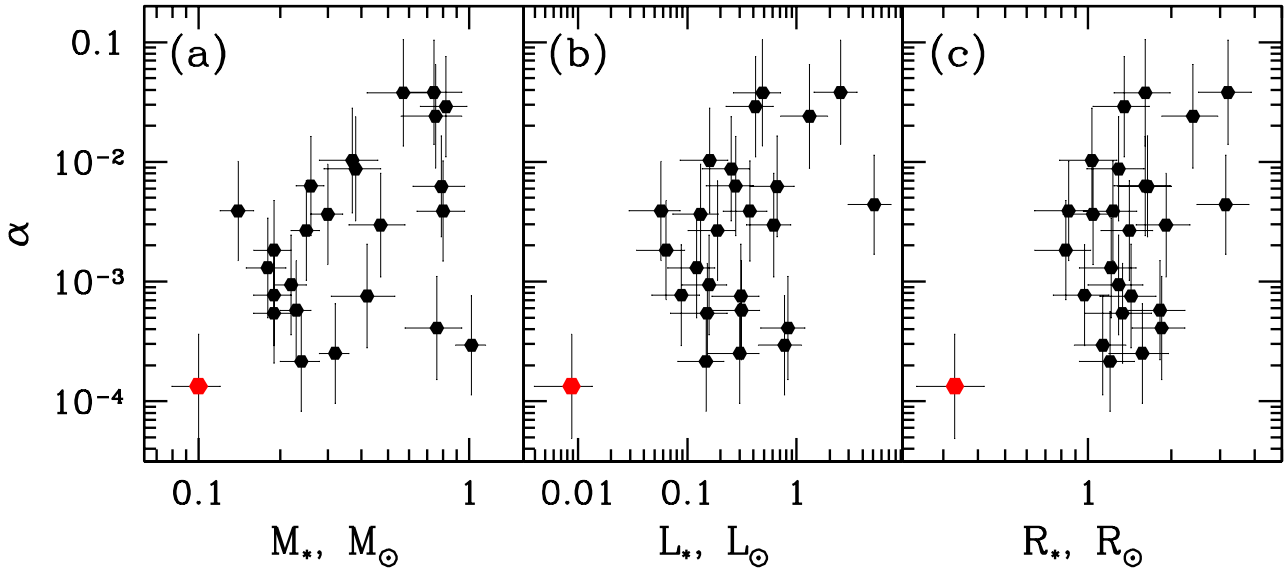}
\caption{Effective viscosity $\alpha$, computed for $T=20$ K, plotted versus stellar parameters: (a) stellar mass $M_\star$, (b) luminosity $L_\star$, and (c) radius $R_\star$. No clear correlations are present in the data, especially when the near-brown dwarf object 2MASS-J16081497-3857145 (red dot) is not included in the sample. 
\label{fig:star_corrs}}
\end{figure*}

In Figure \ref{fig:star_corrs} we examine this possibility, finding no significant correlations between $\alpha$ and the stellar mass, luminosity, or radius. Weak correlations that may be present in the full data set vanish when we remove the brown dwarf-like object 2MASS-J16081497-3857145 (which has very distinct properties and strongly affects covariances between the variables) from the sample. 

This lack of correlation is not surprising from the physical point of view, as one may expect only relatively weak effect of $M_\star$ (e.g. through local shear, proportional to $M_\star^{1/2}$) or $L_\star$ (on which the disk temperature might depend) on the global disk properties.

%%%%%%%%%%%%%%%%%%%%%%%%%%%%%%%%%%%%%%%%%%%%%%%%%%%%%%%%%%%
%%%%%%%%%%%%%%%%%%%%%%%%%%%%%%%%%%%%%%%%%%%%%%%%%%%%%%%%%%%

\section{Discussion}
\label{sect:disc}

%%%%%%%%%%%%%%%%%%%%%%%%%%%%%%%%%%%%%%%%%%%%%%%%%%%%%%%%%%%

Having established a close relation between the mass accretion rate onto the central star $\dot M$ and the inferred value of $\alpha$ on the global scale of the disk, we now seek to understand the implications of this finding. When doing this, it is also important to keep in mind the lack of any significant correlations of $\alpha$ with other obvious characteristics of the system, be it global (like $M_d$ or $\rout$) or local, stellar (e.g. $M_\star$, or $L_\star$). 

There are different ways, in which such a correlation could emerge. First, it may result from various systematic effects related to the measurement of the observables (\S \ref{sect:factors}). Second, there may be a physical reason for the correlation. This would be the case if, for example, some processes related to accretion of gas onto the stellar surface are able to influence the value of $\alpha$ globally, on scales $\sim\rout$ (\S \ref{sect:physical1}). Alternatively, the value of $\alpha$-parameter may depend on some yet unidentified property of the protoplanetary disk, resulting in observed spread, and giving rise to a variation of $\dot M$ (\S \ref{sect:physical2}). Third, the $\alpha$-$\dot M$ relation (\ref{eq:alpha_Mdot}) may simply reflect the way, in which $\dot M$ enters the determination of $\alpha$ in equation (\ref{eq:alpha_expr}), with $\dot M$ being, in fact, largely unrelated to the global disk characteristics. This would be the case if the central $\dot M$ were {\it decoupled} from the global accretion rate set by the disk properties, e.g. as a result of some instability operating in the inner disk, or mass accumulation in a dead zone (\S \ref{sect:forced}). Decoupling would also be natural if the angular momentum transport in protoplanetary disks does not have a diffusive character (\S \ref{sect:non-visc}) and is not characterised by equations (\ref{eq:t_nu})-(\ref{eq:Mdot-Md}).

We now examine each of these possibilities in detail.

%%%%%%%%%%%%%%%%%%%%%%%%%%%
%%%%%%%%%%%%%%%%%%%%%%%%%%%

%%%%%%%%%%%%%%%%%%%%%%%%%%%%%%%%%%%%%%%%%%%%%%%%%%%%%%%%%%%

\subsection{Observational biases}  
\label{sect:factors}

%%%%%%%%%%%%%%%%%%%%%%%%%%%%%%%%%%%%%%%%%%%%%%%%%%%%%%%%%%%

Our calculation of $\alpha$ involves several observables --- $\dot M$, $M_d$, $\rout$ --- and we need to make sure that the origin of the $\alpha$-$\dot M$ correlation is not related to the possible systematic biases in their measurement. We do this next for each of these variables.

%%%%%%%%%%%%%%%%%%%%%%%%%%%%%%%%%%%%%%%%%%%%%%%%%%%%%%%%%%%

\subsubsection{Uncertainty in $\dot M$}  
\label{sect:Mdot_error}

 of stellar $\dot M$ is a challenging task, which was accomplished in \citet{Alcala14,Alcala16} by measuring the UV excess above the stellar photospheric emission. A variety of factors, including the differences between the stellar evolution tracks computed by different groups, contribute to the uncertainty in the subsequent derivation of $\dot M$, which we conservatively adopted to be about 0.4 dex \citep{Alcala16}. However, it is not easy to see how they could enforce a systematic (and not random) correlation (\ref{eq:alpha_Mdot}).

One way to do this might involve the unobserved portion of the accretion luminosity, which could skew the $\dot M$ determination in a systematic way. Indeed, it may be the case that in many systems most of the accretional energy $\approx GM_\star\dot M/R_\star$ is re-emitted in the (high energy) spectral region inaccessible to ground-based instruments. In that case the accretion luminosity measured from the ground would account for only a small fraction of the bolometric accretion flux. If gas accretion is mediated by stellar magnetosphere, which truncates the disk, then one may expect \citep{Calvet_shock} the discrepancy in $\dot M$ determination to correlate with the virial temperature $T_{\rm vir}=(\mu/k)GM_\star/R_\star$ of the gas striking the stellar surface in free fall --- the higher $T_{\rm vir}$ would shift emission from accretion shock to shorter wavelengths and result in a more severe underestimate of $\dot M$. This deviation of $\dot M$ from its true value would then lead to an underestimate of $\alpha$ inferred through equation (\ref{eq:alpha_expr}), perfectly correlated with the biased estimate of $\dot M$.

\begin{figure}
\centering
\includegraphics[width=0.5\textwidth]{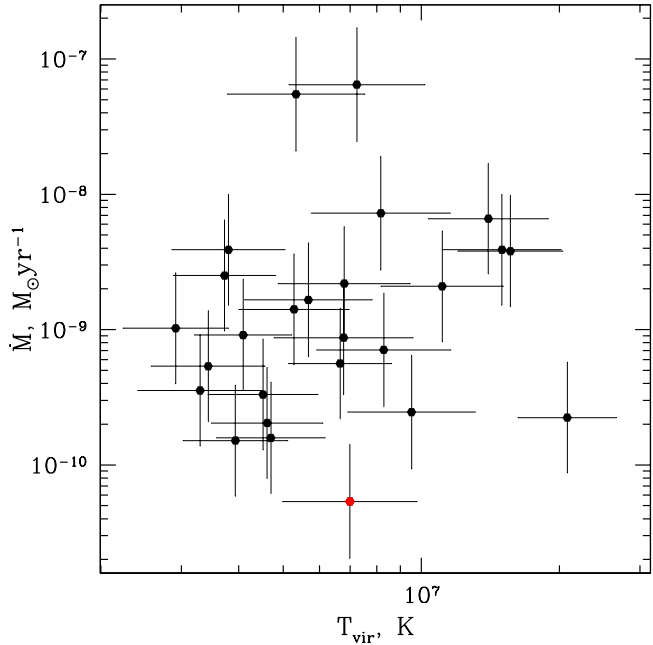}
\caption{
Stellar $\dot M$ plotted vs. the virial temperature at the stellar surface $T_{\rm vir}$. No obvious correlation is seen in the data, demonstrating the lack of biases related to the unobserved fraction of the accretion luminosity.
\label{fig:Mdot_Tvir}}
\end{figure}

To address this issue, in Figure \ref{fig:Mdot_Tvir} we plot $\dot M$ vs. the virial temperature calculated using stellar parameters from Table \ref{table:sample}. One can see no correlation of a kind suggested above, with systems having higher $T_{\rm vir}$ not showing systematically lower values of $\dot M$. The two variables appear completely uncorrelated in our sample. This suggests that the determination of $\dot M$ does not suffer from the bias related to the unobserved accretion luminosity.

%%%%%%%%%%%%%%%%%%%%%%%%%%%%%%%%%%%%%%%%%%%%%%%%%%%%%%%%%%%

\subsubsection{Uncertainty in $\rout$}  
\label{sect:rout_error}

In this work we also implicitly assumed that $\rout$, obtained in \citet{Ansdell} by fitting a Gaussian to the observed intensity pattern, is the true outer disk radius, which encloses its full mass. One may worry that, in fact, this radius corresponds to the sensitivity limit of {\it ALMA} and in reality the disk extends beyond $\rout$, so that both $\rout$ and $M_d$ {\it underestimate} their true values. However, Figure \ref{fig:disk_corrs}c demonstrates that this is not the case: the values of $M_d/r_{\rm out}^2$ proportional to the surface brightness of the outer disk do not cluster around a single value (which could be interpreted as the sensitivity limit of observations) but rather extend over almost two orders of magnitude.

A potentially more serious issue with $\rout$ may arise in systems with different apparent sizes of the gas and dust disks. Evidence for this discrepancy has been found recently in TW Hya \citep{Andrews}, IM Lup \citep{Cleeves}, HD 97048 \citep{Walsh}, with the dust continuum emission being radially more centrally concentrated by a factor of 2-3 than the gaseous disk emitting in $^{12}$CO lines. This has been interpreted as the evidence for the radial inward drift of solids in these disks \citep{Birnstiel}, which decouples radial distributions of the gas and dust surface densities. If this interpretation is correct, then the dust masses would still properly reflect the full disk mass, but the size of the main mass reservoir (gas disk) would be {\it underestimated} by a factor of several. Although this issue should be further explored observationally for our Lupus sample, we believe that it is unlikely to affect our main conclusions for the following reasons.

First, the inferred $\alpha$ depends on $\rout$ rather weakly, e.g. as $\alpha\propto \rout^{1/2}$ if $T=20$ K, see equation (\ref{eq:alpha_expr}). Thus, a possible underestimate of $\rout$ by a factor of $2-3$ would not explain the broad distribution of the inferred values of $\alpha$. Second, it is not clear that the gas disk sizes based on $^{12}$CO measurements represent the radii where most of gas mass is concentrated (which is what the actual $\rout$ should correspond to). Because of the optical thickness of the $^{12}$CO lines, it is generally believed that the CO isotopologues are better tracers of the gas mass distribution than the $^{12}$CO molecule. And the sizes of regions emitting in $^{13}$CO and C$^{18}$O lines tend to be less discrepant with the dust continuum-based radii than the ones based on $^{12}$CO emission  \citep{Schwarz,Cleeves}. This statement seems to hold in our sample too \citep{Ansdell}, based on the disk images obtained using different tracers.

%%%%%%%%%%%%%%%%%%%%%%%%%%%%%%%%%%%%%%%%%%%%%%%%%%%%%%%%%%%

\subsubsection{Uncertainty in $M_d$}  
\label{sect:M_d_error}

Finally, we discus the effect of the uncertainty in the disk mass measurement. \citet{Miotello16} derived more accurate dust continuum-based masses of the disks from the sample of \citet{Ansdell} using detailed radiative transfer calculations of the thermal structure of the disk (instead of assuming a single $T=20$ K as in \citealt{Ansdell}). They found that \citet{Ansdell} systematically overestimate $M_d$ by about a factor of 2 for $M_d\lesssim 10^{-2}M_\odot$. However, this bias would simply {\it uniformly shift} our $\alpha$-$\dot M$ relation, without affecting its scatter or slope. A similar effect would be caused by the possibility of an inward drift of solids (see \S \ref{sect:rout_error}), which tends to decrease the gas-to-dust ratio $\chi$ in the disk region probed by the sub-mm continuum measurements. However, such bias would just shift down (roughly uniformly) the disk mass enclosed within the dust disk radius, without breaking the $\alpha-\dot M$ correlation of increasing the spread of $\alpha$.

Moreover, equation (\ref{eq:alpha_expr}) remains valid even if $\rout$ and $M_d$ do not characterize the full disk: as long as $M_d(<r)$ accounts for the full disk mass enclosed within some radius $r$, their values can be used instead of $M_d$ and $\rout$ for the determination of $\alpha$ via equation (\ref{eq:alpha_expr}). 

Based on this discussion we conclude that observational uncertainties and biases are unlikely causes of the correlation (\ref{eq:alpha_Mdot}) and can hardly account for the full spread in the inferred values of $\alpha$ seen in Figure \ref{fig:hist}. 

%%%%%%%%%%%%%%%%%%%%%%%%%%%%%%%%%%%%%%%%%%%%%%%%%%%%%%%%%%%

\subsection{$\dot M$ setting $\alpha$}  
\label{sect:physical1}

%%%%%%%%%%%%%%%%%%%%%%%%%%%%%%%%%%%%%%%%%%%%%%%%%%%%%%%%%%%

Another possibility for the origin of $\alpha$-$\dot M$ correlation is that the central $\dot M$ has a direct physical effect on $\alpha$. It is difficult to see how this connection can be realized in practice, since $\alpha$ is set by the disk physics on {\it global} scales, while $\dot M$ is a {\it local} property, characterizing the innermost region of the disk.   

%%%%%%%%%%%%%%%%%%%%%%%%%%%
%%%%%%%%%%%%%%%%%%%%%%%%%%%

One possiblity for establishing this connection is if the {\it accretion energy release} at the stellar surface has a direct impact on the value of $\alpha$ on global scales. This may be the case if the value of $\alpha$ depends on the degree of ionization (as may be expected for the non-ideal MRI), and the accretional luminosity plays a major role in determining the ionization balance in the outer disk. If that were the case, one would expect to see a correlation between the global $\alpha$ and the {\it accretion energy flux} $F=GM_\star\dot M/(4\pi R_\star r_{\rm out}^2)$ at $r=\rout$. 

\begin{figure}
\centering
\includegraphics[width=0.5\textwidth]{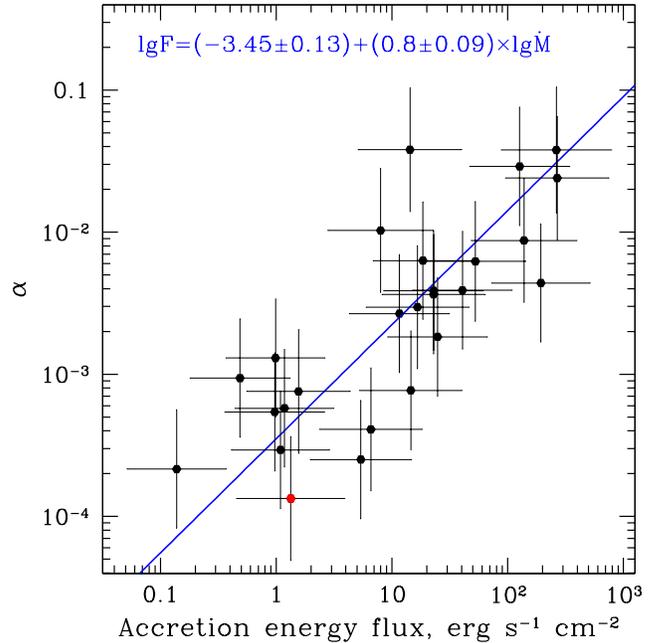}
\caption{
Effective viscosity plotted as a function of the accretion energy flux at the outer radius of the disk, $F=GM_\star\dot M/(4\pi R_\star r_{\rm out}^2)$.
\label{fig:ionize}}
\end{figure}

Figure \ref{fig:ionize} demonstrates that such correlation does indeed exist. However, it shows larger scatter around the best fit line than the correlation in Figure \ref{fig:Mdot_alpha}. This would not be expected if it were $F$ rather than $\dot M$ alone being the real culprit behind the $\alpha$-$\dot M$ correlation. Moreover, it is also unlikely that the spectral range used for inferring $\dot M$ (longward of 310 nm, \citealt{Alcala16}) dominates the ionization balance of the disk. Nor is it clear that the accretion energy release provides major contribution to the flux of ionizing photons impinging on the disk \citep{Glassgold2000}. Furthermore, it is not obvious why this physical mechanism should give rise to an $\alpha(F)$ dependence with a slope so close to unity. 

One final argument against a physical effect of the central $\dot M$ on the global value of $\alpha$ is that the disk with $\alpha(F)\propto F^{0.8}\propto r^{-1.6}$ dependence, suggested by the best fit in Figure \ref{fig:ionize}, should have a rather unusual structure. Indeed, inside $\rout$ the disk should converge to a constant $\dot M$ structure, meaning that $\Sigma\propto\dot M/\nu\propto r^{0.1}T^{-1}$, where we used equation (\ref{eq:alpha}) for $\nu$ and took $\alpha\propto r^{-1.6}$. Since the disk temperature $T$ does not increase with $r$, this would mean that $\Sigma$ should be an {\it increasing} function of $r$. This conclusion is hardly compatible with our understanding of the protoplanetary disk structure.

For all these reasons we do not find the direct physical effect of stellar $\dot M$ on $\alpha$ to be a plausible explanation of the $\alpha$-$\dot M$ correlation.

%%%%%%%%%%%%%%%%%%%%%%%%%%%%%%%%%%%%%%%%%%%%%%%%%%%%%%%%%%%

\subsection{$\alpha$ setting $\dot M$}  
\label{sect:physical2}

%%%%%%%%%%%%%%%%%%%%%%%%%%%%%%%%%%%%%%%%%%%%%%%%%%%%%%%%%%%

Physical connection between $\alpha$ and $\dot M$ may also emerge in the direction opposite to that considered in \S \ref{sect:physical1}, with $\alpha$ directly affecting $\dot M$. Such connection is rather natural in light of the equation (\ref{eq:alpha_expr}). However, in the conventional picture $\alpha$ has a unique value, which is incompatible with the distribution shown in Figure \ref{fig:hist}. Thus, in the observed sample the broad range of $\alpha$ has to be caused by some additional environmental parameter, which controls angular momentum transport and allows $\alpha$ to vary over almost three orders of magnitude. And then $\alpha$-$\dot M$ correlation would naturally emerge from equation (\ref{eq:alpha_expr}), with the distribution of $\alpha$ directly translating into the broad range of the $\dot M$ values. 

The hidden parameter controlling $\alpha$ cannot be one of the global disk variables --- $M_d$, $\rout$, global surface density --- as Figure \ref{fig:disk_corrs} shows no correlation of $\alpha$ with them. Cooling time of the disk, which directly depends on these global disk characteristics, also cannot be the controlling parameter. This likely excludes the vertical shear instability \citep{Urpin}, which sensitively depends on the local cooling time \citep{Stoll,Lin}, from being a candidate for driving the viscous evolution of the protoplanetary disks.   

%%%%%%%%%%%%%%%%%%%%%%%%%%%
%%%%%%%%%%%%%%%%%%%%%%%%%%%

At the same time, there is a variety of possible controlling parameters if transport in the disk is effected by the non-ideal MRI. They include (but are not limited to) ionization fraction on scales $\sim \rout$ \citep{Jin,Fleming,Bai11}, strength of the magnetic field in the disk \citep{Bai11}, or its geometry \citep{Simon1,Simon2}. All of these physical characteristics are difficult to determine observationally at the moment.  Nevertheless, we believe that this way of producing $\alpha$-$\dot M$ correlation is more plausible than the one outlined in \S \ref{sect:physical1}.

%%%%%%%%%%%%%%%%%%%%%%%%%%%%%%%%%%%%%%%%%%%%%%%%%%%%%%%%%%%

\subsection{Decoupling of stellar $\dot M$ from the global mass accretion rate}  
\label{sect:forced}

%%%%%%%%%%%%%%%%%%%%%%%%%%%%%%%%%%%%%%%%%%%%%%%%%%%%%%%%%%%

A correlation between the inferred $\alpha$ and $\dot M$ would also naturally emerge if the $\dot M$ measured through stellar accretional luminosity is, in fact, {\it unrelated} to the global mass accretion rate $M_d/t_\nu$. 

In the standard viscous disk theory the two should be equal, as demonstrated by the equation (\ref{eq:Mdot-Md}). However, if stellar $\dot M$ is somehow {\it decoupled} from the global accretion rate, then equation (\ref{eq:alpha_expr}) would naturally result in a strong linear correlation between the {\it inferred} $\alpha$ (unrelated to the real global $\alpha$) and $\dot M$. This would be true even if the real $\alpha$ set by the physics of the angular momentum transport on scales $\sim \rout$ takes on a unique value. Errors in measuring $\dot M$ could lead to this situation, but they would need to be very dramatic (potentially exceeding two orders of magnitude), which is unlikely, as we showed in \S \ref{sect:factors}.

A decoupling between $\dot M$ and $M_d/t_\nu$ (by more than two orders of magnitude, to explain the range of inferred $\alpha$) requires a modification of the simple picture of viscous disk accretion. It can arise, for example, if some instability operates in the inner regions of protoplanetary disks, dramatically modulating local $\dot M$ compared to its global value set on scales $\sim \rout$. The characteristic timescale for such variability should be  substantial for it to have escaped detection until now. One may suspect FUor and EXor outbursts \citep{Audard} to be the known realizations of such an instability. However, one would then expect the distribution of $\alpha$ to be {\it bimodal}, with most disks being in quiescence and having low $\alpha$, and a small population of disks undergoing an outburst and having high inferred $\alpha$ \citep{Audard}.

Another way of decoupling stellar $\dot M$ from the global accretion rate is if the viscous mass flow towards the star accumulates in a substantial mass reservoir at some intermediate radii, e.g. in a dead zone \citep{GammieDZ}. This reservoir should be able to accumulate large amounts of mass, comparable to the total disk mass at the start of its evolution. This may be difficult to realize on timescales comparable to the disk lifetime (Myrs), necessitating periodic deposition of mass from the reservoir onto the star, and making this scenario similar to the aforementioned instability in the inner disk.

Alternatively, gas reaching the inner regions of the disk may be {\it lost} in a wind (photoevaporative or MHD-driven) or consumed by vigorously accreting planets. If that were the case, then in many systems the mass loss rate should be matching the global accretion rate set on large scales, with only a small amount of mass reaching the star. The wind is likely to also affect the angular momentum budget of the disk in a non-trivial manner, a possibility that we consider next.

%%%%%%%%%%%%%%%%%%%%%%%%%%%%%%%%%%%%%%%%%%%%%%%%%%%%%%%%%%%

\subsection{Non-viscous evolution of the protoplanetary disks}
\label{sect:non-visc}

%%%%%%%%%%%%%%%%%%%%%%%%%%%%%%%%%%%%%%%%%%%%%%%%%%%%%%%%%%%

One final, very intriguing possibility, is that the angular momentum and mass transport in the protoplanetary disks has a non-viscous (non-diffusive) character. In this case equations (\ref{eq:t_nu})-(\ref{eq:Mdot-Md}) do not hold, and $\alpha$-$\dot M$ correlation emerges simply as a consequence of calculating $\alpha$ through equation (\ref{eq:alpha}), with no real physical meaning for $\alpha$. Also, stellar $\dot M$ may have little to do with with the global disk parameters, although the work of \citet{Manara} does show evidence for a correlation between $M_d$ and $\dot M$ (which is not obvious in our sample of resolved disks).

Such non-viscous transport may be effected in protoplanetary disks by the magnetically-controlled winds \citep{BP,Konigl}. Outflows from the disks of YSOs are a well studied observational phenomenon \citep{Frank}. Recently self-consistent launching of the magnetocentrifugal winds has been observed in simulations of magnetized accretion disks \citep{Suzuki,Bai}, adding support to this possibility. 

Another potential driver of the non-diffusive evolution of the protoplanetary disks could be the density waves excited by massive perturbers, e.g. planets or stellar companions \citep{GR01,Rafikov02,Rafikov16,Dong}. Global spiral waves have been observed recently in several protoplanetary disks in scattered light, e.g. in SAO 206462 \citep{Garufi}, MWC 758 \citep{Benisty}, HD 100453 \citep{Wagner}, etc.

We believe that in light of the perceived difficulty of the known local turbulent transport mechanisms to drive the  protoplanetary disk evolution on Myr timescales \citep{Turner}, the non-diffusive mechanisms for driving disk evolution should be considered very seriously. Our work may thus provide strong indirect evidence in favor of this possibility.  

%%%%%%%%%%%%%%%%%%%%%%%%%%%%%%%%%%%%%%%%%%%%%%%%%%%%%%%%%%%

\subsection{Comparison with previous studies}
\label{sect:compare}

%%%%%%%%%%%%%%%%%%%%%%%%%%%%%%%%%%%%%%%%%%%%%%%%%%%%%%%%%%%

There has been a handful of studies trying to understand viscous evolution of the protoplanetary disks based on observational data. Using the mean properties of a large sample of the protoplanetary disks \citet{Hartmann} have concluded that their effective viscosity should be narrowly clustered around $\alpha\approx 10^{-2}$. In this work we determine $\alpha$ for {\it individual objects}, and find a much broader distribution of $\alpha$, extending over more than two orders of magnitude (Figure \ref{fig:hist}). This difference suggests that care should be taken when making inferences based on the averaged properties of the sample.

Some studies of viscous evolution of the protoplanetary disks have tried to verify the equation (\ref{eq:Mdot-Md}) with $t_\nu$ set equal to the age of the system $t_\star$ \citep{Hartmann,Jones,Manara}. Identification of $t_\nu$ with the age of the central star is a valid procedure as long as $t_\star$ exceeds the viscous time of the disk at its initial radius $r_0$. This assumption is similar to our implicit assumption that the current disk size exceeds its initial size, $\rout\gtrsim r_0$, so that viscous evolution enters the self-similar regime and the memory of initial conditions gets erased. However, the problem with using this metric of disk evolution is that the determination of ages of the young stars is notoriously difficult \citep{Soderblom}.

\begin{figure}
\centering
\includegraphics[width=0.5\textwidth]{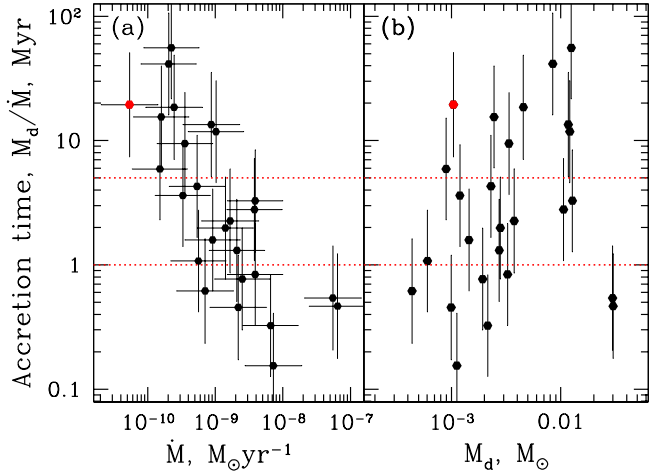}
\caption{
Characteristic accretion time $M_d/\dot M$ shown as a function of (a) $\dot M$ and (b) $M_d$. Dotted lines illustrate the approximate upper and lower age limits for the objects in Lupus \citep{Alcala14}.
\label{fig:acc_time}}
\end{figure}

Despite this drawback \citet{Hartmann} and \citet{Jones} have tried to verify that $M_d/\dot M\sim t_\star$ (up to a constant factor of order unity) using observational data. Both studies found significant deviations (up to two orders of magnitude) from this simple relation. In Figure \ref{fig:acc_time} we show the characteristic accretion time for the objects in our sample, plotted against $\dot M$ and $M_d$. Dotted lines show the range of ages for the Lupus objects in our sample \citep{Alcala16} (we do not attempt to use uncertain ages of the individual objects). It is clear that accretion times of many objects fall outside this range, by more than an order of magnitude in some cases. This agrees with the conclusions of \citet{Hartmann} and \citet{Jones}.

In Figure \ref{fig:acc_time}a one can also see a strong anti-correlation between $M_d/\dot M$ and $\dot M$, with a much weaker statistical connection for the $M_d$ (see Figure \ref{fig:acc_time}b). This demonstrates the key role of $\dot M$ for the accretion time, just as we found for $\alpha$ in \S \ref{sect:mdot}. 

\citet{Jones} were not able to account for the discrepancy between $M_d/\dot M$ and $t_\star$ even using sophisticated disk models including the effects of dead zones, photoevaporation, planet formation, etc. Instead, they concluded that it follows from the systematic errors in the determination of $M_d$. However, we find that the systematic underestimate (or overestimate) of $M_d$ would result in a uniform overestimate (underestimate) of $\alpha$, but would not explain the emergence of the $\alpha$-$\dot M$ correlation. Our results suggest that the discrepancy between $M_d/\dot M$ and $t_\star$ is more likely to be caused by the decoupling of the central $\dot M$ from the global mass accretion rate $M_d/t_\nu$ computed based on the standard theory of viscous disk evolution, as described in \S \ref{sect:forced}-\ref{sect:non-visc}. 

We also note that the scenario, in which $\alpha$ is controlled by a yet unidentified variable (\S \ref{sect:physical2}), is not expected to produce significant deviations from $M_d/\dot M\sim t_\star$ relation (which is insensitive to $\alpha$ in the self-similar regime). This may argue against this scenario for the protoplanetary disks, although more work is certainly needed to resolve this question. 

%%%%%%%%%%%%%%%%%%%%%%%%%%%
%%%%%%%%%%%%%%%%%%%%%%%%%%%

Our approach bypasses the issue of uncertain stellar ages by simply ignoring them altogether. Instead, we use {\it spatial information} to gain insight on the physical mechanisms responsible for the angular momentum transport in the protoplanetary disks by measuring $\alpha$. Past efforts \citep{Hartmann,Jones} did not have the ability to do that because they lacked the accurate information on the sizes of disks in individual objects. Thus, our work represents an independent way of testing the theory of viscous evolution of the protoplanetary disks.

%%%%%%%%%%%%%%%%%%%%%%%%%%%%%%%%%%%%%%%%%%%%%%%%%%%%%%%%%%%
%%%%%%%%%%%%%%%%%%%%%%%%%%%%%%%%%%%%%%%%%%%%%%%%%%%%%%%%%%%

\section{Summary}  
\label{sect:sum}

%%%%%%%%%%%%%%%%%%%%%%%%%%%%%%%%%%%%%%%%%%%%%%%%%%%%%%%%%%%

In this work we explored viscous evolution of the protoplanetary disks. Using observational sample of 26 disks resolved with {\it ALMA} with measured masses (based on sub-mm continuum) and central accretion rates we derived the values of the dimensionless viscosity parameter $\alpha$, with the goal of constraining the mechanism of the angular momentum transport in the disk. Our findings can be summarized as follows.

\begin{itemize}

\item The distribution of inferred values of $\alpha$ extends over more than two orders of magnitude, from $10^{-4}$ to $0.04$, with no obvious preferred value inside this interval.

\item We found no correlation of $\alpha$ with either the global disk parameters --- mass, size, surface density --- or stellar parameters --- luminosity, mass, radius.

\item The main finding of this work is the discovery of a strong linear correlation between $\alpha$ and central mass accretion rate $\dot M$, which is robust with regard to the thermodynamic assumptions about the disk. This correlation persists even if we use the CO-based gas masses for computing $\alpha$, and holds not only for $\alpha$ but also for the dimensional kinematic viscosity $\nu$ on global scales.

\end{itemize}

These results suggest that a simple picture, in which viscous evolution of the protoplanetary disks is driven by a physical process (e.g. MRI) with a single, well-defined value of $\alpha$, is too simplistic and must be modified. We find that observational errors and biases cannot account for the observed $\alpha$-$\dot M$ correlation, and seek other explanations. We find it unlikely that gas accretion onto the stellar surface can have a direct effect on $\alpha$ (e.g. through the accretional energy release) on scales of order the disk size (tens to hundreds of AU). We propose three other possibilities for explaining $\alpha$-$\dot M$ correlation, which effectively assume that either $\alpha$ or $\dot M$ are decoupled from the global characteristics (mass, size) of the disk. In that case equation (\ref{eq:alpha}) naturally leads to a linear relation between $\alpha$ and $\dot M$. These possibilities are as follows.

\begin{itemize}

\item The value of $\alpha$ in every disk is controlled by some yet unobserved variable, variation of which is responsible for the broad range of $\alpha$. This, in turn, is the main cause of the variation of $\dot M$. In the case of accretion driven by the (non-ideal) MRI the role of such control parameter may be played by the disk ionization, as well as the strength or geometry of the magnetic field in the disk.

\item Stellar $\dot M$ may be decoupled from the global mass accretion rate by some instability operating in the inner disk, or mass accumulation in a dead zone, or a wind with high mass loss rate. In this case the inferred values of $\alpha$ do not characterize the global disk evolution.

\item Finally, disk evolution may have a non-diffusive (non-viscous) character, in which case the inferred $\alpha$ has no physical meaning. This may be the case if mass accretion in protoplanetary disks is driven by e.g. magnetocentrifugal winds or spiral density waves.

\end{itemize}

Future work aimed at expanding the sample of resolved protoplanetary disks with well measured masses and accretion rates will help us to identify the physical reason behind the observed $\alpha$-$\dot M$ correlation.
\\

\acknowledgements

I am indebted to Megan Ansdell and Juan Manuel Alcal\'a for sharing their data with me, to Carlo Felice Manara and Eugene Churazov for useful discussions, and to Ruobing Dong for insightful comments on the manuscript. Financial support for this study has been provided by the NSF via grant AST-1409524 and NASA via grant 15-XRP15-2-0139.

%%%%%%%%%%%%%%%%%%%%%%%%%%%%%%%%%%%%%%%%%%%%%%%%%%%%%%%%%%%
%%%%%%%%%%%%%%%%%%%%%%%%%%%%%%%%%%%%%%%%%%%%%%%%%%%%%%%%%%%

\bibliographystyle{apj}
\bibliography{references}

%%%%%%%%%%%%%%%%%%%%%%%%%%%
\begin{sidewaystable}
%\begin{table*}
\begin{threeparttable}
\caption{Observational sample}
\begin{tabular}{ccccccccc}
    \hline \hline\\ 
Name & $L_\star$, $L_\odot$ & $R_\star$, $R_\odot$ & $M_\star$, $M_\odot$ & $\lg [\dot M/(M_\odot \mbox{yr}^{-1})]$ & SMA, arcsec & $M_{\rm dust}$, $M_\oplus$ & $M_g [\min(M_g), \max(M_g)]$, $M_J$ & $d$, pc\\ 
\\ 
\hline
Sz65	 &		 $0.8318\pm  0.3623$ &  $ 1.84\pm	0.40$  &   $0.76\pm	0.18$ &    $-9.61\pm   0.42$ & $0.171 \pm 0.002$ & $15.1559 \pm 0.0752$ & 0.64 [0.2, 1.5] &	150\\   
Sz68 &			 $5.1286 \pm 2.1919 $ &  $3.14\pm	0.67$  &   $2.13\pm	0.33$ &    $-8.42 \pm  0.41$ &  $0.159\pm 0.002$ & $35.3387\pm 0.1081$ & 0.68 [0.2, 1.5] &	150\\  
Sz69 &	$0.0880\pm 0.0410$ & $0.97\pm 0.22$ & $0.19\pm	0.03$ &	$-9.48\pm 0.41$ &	$0.092\pm 0.012$ & $3.9858 \pm 0.0658$ & 0.034 [0.018, 0.07] &	150\\ 
Sz71 &	$0.3090\pm	0.1420$ & $1.43\pm 0.33$ & $0.42\pm	0.11$ & $-9.06\pm 0.42 $ & $0.558\pm   0.003$ & $39.0213\pm  0.1481$ & 0.096 [0.07, 0.3] &	150\\ 
Sz72 &	$0.2520\pm	0.1160$ &  $ 1.29\pm	0.30$ &   $ 0.38\pm	0.09$ &   $ -8.66\pm   0.42 $ &  $0.094\pm   0.012 $ & $3.3137 \pm 0.0658$ & - &	150\\ 
Sz73 & $0.4190\pm	0.1930$ &  $1.35\pm	0.31$ &    $0.82\pm	0.16$ &    $-8.18\pm   0.41$ & $0.245\pm   0.01$ &    $7.1514\pm  0.1293$ & - &	150\\ 
Sz83 &	$1.3130\pm	0.6050$ &   $2.39\pm	0.55$ &   $ 0.75\pm	0.19$ &    $-7.19\pm   0.42 $ & $ 0.379\pm   0.001$ & $100.3265 \pm 0.1692$ & 1.5 [0.48, 4.0] &	150\\ 
Sz84 (td) &	$0.1220\pm	0.0560$ &   $1.21\pm	0.28$ &    $0.18\pm	0.03$ &    $-9.27\pm   0.41$ & $0.392 \pm  0.006$ &    $7.6708 \pm 0.094$ & 0.11 [0.06, 0.22] &	150\\ 
Sz88A &	$0.4880\pm	0.2250$ &   $1.61\pm	0.37$ &    $0.57\pm	0.15$ &    $-8.14 \pm  0.42$ & $0.102\pm   0.02$ &     $3.7351\pm  0.1253$ & - &	200\\  
Sz90	 &		 $0.6607 \pm 0.2845$ &  $ 1.64\pm	0.36 $  & $0.79\pm	0.17$ &    $-8.68 \pm  0.41$ & $0.143 \pm 0.011$ & $9.1205 \pm 0.1922$	 & 0.056 [0.035, 0.1] &	200\\  
Sz98 &			 $2.5119 \pm 1.0755$ &   $3.20\pm	0.69$ & $ 0.74\pm 0.20$ & $-7.26 \pm  0.42$ & $0.974 \pm  0.006$ & $99.1394\pm  0.5933$ & 0.066 [0.04, 0.1] &	200\\ 
Sz103 &	$0.1880\pm	0.0870$ &   $1.41\pm	0.30$ &    $0.25\pm	0.03$ &    $-9.04\pm   0.41$ &  $0.122\pm   0.012$  &    $4.8214\pm  0.117$ & - &	200\\ 
Sz108B	 &		$ 0.1514\pm  0.0813$ &   $1.33\pm	0.36$ &   $ 0.19\pm	0.03$ & $-9.45\pm  0.41$ &  $0.236\pm  0.005$ & $11.1845 \pm 0.1421$ & 0.65	[0.2, 1.5] &	200\\   
Sz110 &	$0.2760\pm	0.1270$ &   $1.61\pm	0.37$ &    $0.26\pm	0.03$ &    $-8.60\pm   0.41$ & $0.153\pm   0.009$  &   $6.4341\pm  0.1212$ & - &	200\\  
Sz113 &	$0.0640\pm	0.0300$ &   $0.83\pm	0.19$ &    $0.19\pm	0.03$ &    $-8.85\pm   0.41$ & $0.118\pm   0.007$ &     $9.3378\pm  0.1128$ & - &	200\\ 
Sz114 &	$0.3120\pm	0.1440$ &   $1.82\pm	0.42$ &    $0.23\pm	0.03$ &    $-8.99\pm   0.41$ & $0.342\pm   0.002$ & $40.28 \pm   0.1713$ &  0.096 [0.065, 0.28] & 200\\  
Sz129 &	$ 0.3715 \pm 0.1600$ &   $1.23\pm	0.27$ &    $0.80\pm	0.16$ &   $ -8.41 \pm  0.41$ & $0.458\pm   0.002$ &  $ 42.5653 \pm 0.1222$ & 0.046 [0.03, 0.09] & 150\\   
Sz130 &	$0.1600\pm	0.0740$ &   $1.03\pm	0.24$ &    $0.37\pm	0.09$ &    $-9.15\pm   0.42$ & $0.246\pm   0.028$ & $1.4547 \pm 0.0823$ & 0.036 [0.011, 0.05] &	150\\ 
Sz131 &	$0.1318\pm  0.0583$ &   $1.04\pm	0.23$ &   $ 0.30\pm	0.04$ &    $-9.25 \pm  0.41$ & $ 0.116 \pm  0.018$ &     $2.0141 \pm 0.0682$ & - & 150\\ 
MYLup (td)	 & $0.7762 \pm 0.3315$ &   $1.13\pm	0.24$ &    $1.02\pm	0.13$ &    $-9.65\pm   0.41$ & $0.593 \pm  0.003$ & $41.5524 \pm 0.1786$ & 0.083 [0.05, 0.21] & 150\\ 
SSTc2d-J154508.9-341734 & $0.0575\pm  0.0283$ & $0.85\pm 0.21$ &    $0.14\pm	0.02$ & $-8.41\pm 0.41$ & $0.173 \pm 0.005$ & $10.874\pm 0.1175$ & 0.77 [0.25, 2.0] & 150\\   
SSTc2d-J160002.4-422216	 & $0.1479\pm  0.0666$ & $1.20\pm 0.27$ & $0.24\pm	0.04$ & $-9.69\pm  0.41$ & $0.749\pm 0.004$ & $28.1662\pm 0.1481$ & 0.14 [0.11, 0.7] & 150 \\  
2MASSJ-16085324-3914401 & $0.3020 \pm 0.1477$ & $1.57\pm 0.38$ & $0.32\pm 0.04$  & $-9.80 \pm  0.41$ & $0.08 \pm 0.009$ & $8.1763\pm 0.117$ & 0.034 [0.02, 0.07] &	200\\  
SSTc2d-J161029.6-392215 & $ 0.1585\pm 0.0698$ & $1.29\pm 0.29$ & $0.22\pm 0.03$ & $-9.82 \pm  0.41$ &  $ 0.238 \pm  0.021$ & $2.9831\pm 0.1462$ & 0.16 [0.1, 0.56] & 200\\   
SSTc2d-J161243.8-381503	 & $0.6166\pm  0.2691$ &   $1.91\pm	0.42$ &    $0.47\pm	0.11$ &    $-8.78 \pm  0.42$ &  $0.162 \pm  0.008$  &   $12.4838 \pm 0.2047$ & - &	200\\  
2MASS-J16081497-3857145	 & $0.0087\pm 0.0047$ & $0.33\pm 0.09$ & $0.10\pm 0.02$ &  $-10.27\pm 0.42$ & $0.114\pm 0.019$ & $3.4761\pm 0.1253$ & 0.022 [0.01, 0.045] & 200 \\  
\hline
\end{tabular}
\begin{tablenotes}
\item {\bf Notes} For each object (Name) we list the luminosity $L_\star$, radius $R_\star$, mass $M_\star$, and accretion rate $\dot M$ of its parent star, the outer semi-major axis (SMA) and dust and gas masses ($M_{\rm dust}$ and $M_g$, the latter with upper and lower limits) of the disk, as well as the distance to the object $d$. Data come from \citet{Alcala14,Alcala16}, \citet{Ansdell}, \citet{Miotello16}. (td) near the object name indicates a transitional disk.
\end{tablenotes}
\label{table:sample}
\end{threeparttable}
%\end{table*}
\end{sidewaystable}
%%%%%%%%%%%%%%%%%%%%%%%%%%%

%%%%%%%%%%%%%%%%%%%%%%%%%%%
\begin{table*}
\begin{threeparttable}
\caption{Statistical characteristics of the data}
\begin{tabular}{ccccc}
    \hline \hline\\
    Variables & Figure & $\rho$ & $r_s$ & $p$-value \\ 
    \\
    \hline
    $\dot M$-$M_d$ & \ref{fig:sample}a & 0.446 & 0.3 & 0.137  \\ 
    %\hline
    $r_{\rm out}$-$M_d$ & \ref{fig:sample}b & 0.7 & 0.631 & $5.5\times 10^{-4}$  \\ 
    %\hline
    $\alpha$-$M_d$ & \ref{fig:disk_corrs}a & 0.004 & -0.093 & 0.653  \\ 
    %\hline
    $\alpha$-$r_{\rm out}$ & \ref{fig:disk_corrs}b & -0.01 & -0.045 & 0.828  \\ 
    %\hline
    $\alpha$-$M_d r_{\rm out}^{-2}$ & \ref{fig:disk_corrs}c & 0.02 & 0.054 & 0.792  \\ 
    %\hline
    $\alpha$-$(M_\star r_{\rm out})^{1/2}M_d^{-1}$ & \ref{fig:disk_corrs}d & 0.18 & 0.209 & 0.306  \\ 
    %\hline
    $\alpha$-$M_d$ & \ref{fig:Mdisk_alpha_T}a & 0.036 & -0.084 & 0.684  \\ 
    %\hline
    $\alpha$-$M_d$ & \ref{fig:Mdisk_alpha_T}b & 0.028 & -0.082 & 0.689  \\ 
    %\hline
    $\alpha$-$\dot M$ & \ref{fig:Mdot_alpha} & 0.877 & 0.868 & $9.4\times 10^{-9}$ \\ 
    %\hline
    $\alpha $-$\dot M$ & \ref{fig:Mdot_alpha_T}a & 0.866 & 0.854 & $2.8\times 10^{-8}$  \\ 
    %\hline
    $\alpha$-$\dot M$ & \ref{fig:Mdot_alpha_T}b &  0.868 & 0.858, & $2.1\times 10^{-8}$ \\ 
    %\hline
    $\alpha $-$\dot M$ & \ref{fig:alpha_gas}& 0.808 & 0.743 & $4.1\times 10^{-4}$ \\ 
    %\hline
    $\nu$-$\dot M$ & \ref{fig:disk_corrs_nu}a & 0.841 & 0.767 & $4.9\times 10^{-6}$ \\ 
    %\hline
    $\nu$-$M_d$ & \ref{fig:disk_corrs_nu}b & 0.304 & 0.119 & 0.564  \\ 
    %\hline
    $\nu$-$r_{\rm out}$ & \ref{fig:disk_corrs_nu}c & 0.517 & 0.421 & 0.032 \\ 
    %\hline 
    $\alpha $-$ M_\star$ & \ref{fig:star_corrs}a & 0.439 (0.354) & 0.414 (0.341) & 0.035 (0.095) \\ 
    %\hline
     $\alpha$-$L_\star$ & \ref{fig:star_corrs}b & 0.448 (0.326) & 0.373 (0.295) & 0.061 (0.153) \\ 
    %\hline
     $\alpha$-$R_\star$ & \ref{fig:star_corrs}c & 0.45 (0.312) & 0.335 (0.252) & 0.094 (0.223) \\ 
    %\hline
    $\dot M$-$T_{\rm vir}$ & \ref{fig:Mdot_Tvir} & 0.177 & 0.235 & 0.247  \\ 
    %\hline
     $\alpha$-$F_{\rm acc}$ & \ref{fig:ionize} & 0.79 & 0.768 & $4.6\times 10^{-6}$ \\ 
    \hline   
\end{tabular}
\begin{tablenotes}
\item {\bf Notes}.  Variables: combination of physical parameters for which the correlation is assessed; Figure: number of the Figure in which this correlation is illustrated; $\rho$: Pearson correlation; $r_s$: Spearman's rank correlation coefficient; $p$-value: probability of a null hypothesis that the two variables are completely uncorrelated. Values in parentheses correspond to the sample with the near-brown dwarf object 2MASS-J16081497-3857145 excluded.
\end{tablenotes}
\label{table:stats}
\end{threeparttable}
\end{table*}
%%%%%%%%%%%%%%%%%%%%%%%%%%%

\end{document}